\documentclass{iaus}
\usepackage{graphicx}

\title[Exoplanets at high spectral resolution] %% give here short title %%
{Exoplanet-atmospheres\\ at high spectral resolution:\\ A CRIRES survey of hot-Jupiters}

\author[Ignas Snellen et al.]   %% give here short author list %%
{Ignas Snellen$^1$, Remco de Kok$^2$, Ernst de Mooij$^1$, Matteo Brogi$^1$,\\ Bas Nefs$^1$ 
\and Simon Albrecht$^3$}

\affiliation{$^1$Leiden Observatory, Leiden University, Postbus 9513, 2300 RA, Leiden, the Netherlands \\ email: {\tt snellen@strw.leidenuniv.nl} \\[\affilskip]
$^2$SRON, Sorbonnelaan 2, 3584 CA Utrecht, The Netherlands\\[\affilskip]
$^3$Department of Physics, and Kavli Institute for Astrophysics and Space Research, Massachusetts Institute of Technology, Cambridge, Massachusetts 02139, USA
}

\pubyear{2010}
\volume{276}  %% insert here IAU Symposium No.
\pagerange{119--123}
\jname{The Astrophysics of Planetary Systems: Formation, Structure, and Dynamical Evolution}
\editors{A. Sozzetti, Mario G. Lattanzi, \& A.P. Boss, eds.}
\begin{document}

\maketitle

\begin{abstract}
Recently, we presented the detection of carbon monoxide in the transmission spectrum of  extrasolar planet HD209458b, using CRIRES, the Cryogenic high-resolution Infrared Echelle Spectrograph at ESO's Very Large Telescope (VLT). The high spectral resolution observations (R=100,000) provide a wealth of information on the planet's orbit, mass, composition, and even on its atmospheric dynamics.  The new observational strategy and data analysis techniques open up a whole world of opportunities. We therefore started an ESO large program using CRIRES to explore these, targeting both transiting and non-transiting planets in carbon monoxide, water vapour, and methane. Observations of the latter molecule will also serve as a test-bed for METIS, the proposed mid-infrared imager and spectrograph  for the European Extremely Large Telescope.

%% add here a maximum of 10 keywords, to be taken form the file <Keywords.txt>
\end{abstract}

\firstsection % if your document starts with a section,
              % remove some space above using this command.
              
\section{Introduction}              

A planet which crosses the disk of its host star allows its atmospheric properties to be studied in three ways: 1) by transmission spectroscopy, when during a transit starlight filters through a planet's atmosphere leaving a spectral imprint of its atmospheric constituents; 2) by eclipse photometry, when the planet moves behind the star the planet light is temporary blocked, and the contribution from the planet can be determined; and 3) by orbital phase variations, since thoughout the orbit we see varying contributions form the (warm) dayside and (cooler) nightside of the planet, which is seen as minute changes in brightness of the system as function of orbital phase. 

Until recently, all the fascinating discoveries using these techniques came from the Hubble and Spitzer space observatories. Transmission spectroscopy and secondary eclipse measurements have yielded detections of absorption signatures from hydrogen, oxygen and carbon atoms in the ultraviolet (Vidal-Madjar et al. 2003;2004), sodium in the optical (Charbonneau et al. 2002), and of broadband molecular signatures in the near- and mid-infrared from water, methane, carbon-monoxide and carbon-dioxide (e.g. Tinetti et al. 2007; Swain et al. 2008; Grillmair et al. 2008; Swain et al. 2009a,b; Beaulieu et al. 2009). The strengths of molecular bands at different wavelengths constrain atmospheric temperature profiles, indicating that some of the hottest planets contain a thermal inversion layer (e.g. Burrows et al. 2007). In addition, orbital phase variations have been detected in the mid-infrared showing the day-to-nightside temperature distributions (Knutson et al. 2007), work that has been recently extended to the optical by CoRoT and Kepler (Snellen et al. 2009; Burucki et al. 2009). 

For long, ground-based observations did not play any significant role, and it was thought that the disturbing influence of the Earth atmosphere, as emphasized by many unsuccessful attempts, would prevent contributions from ground-based instrumentation to this exciting field of research. 
However, over recent years, novel observation and data-analysis techniques have been in development, utilizing the many superior qualities of ground- based instrumentation (such as collective power and spectral resolution) over the space telescopes. This has resulted in a string of ground-based detection of transmission features (e.g. Redfield et al. 2008; Snellen et al. 2008; Sing et al. 2010), and optical and near-infrared eclipse measurements (e.g. Sing \& Lopez-Morales 2009; de Mooij \& Snellen 2009; Croll et al. 2010).

\section{Molecular absorption in a planet's atmosphere at high spectral resolution}

Broad-band space-based observations contain always a level of ambiguity, because molecular signatures overlap, and because high-temperature line lists for some molecular species (e.g. methane) are known to be incomplete. These issues can be circumvented by observing at high spectral resolution and thereby unambiguously identifying the individual molecular absorption lines. We therefore observed one transit of HD209458b with the Cryogenic Infrared Echelle Spectrograph (CRIRES) on the VLT,  targeting the rotation-vibration band of carbon monoxide between 2.29 and 2.34 $\mu$m at a spectal resolution of R=100,000. By combining the signal of $\sim$50 CO absorption lines in the observed wavelength regime we detected a clear signal from the planet's atmosphere (\cite[Snellen et al.(2010)]{} $-$ S10).
The change in the radial component of the planet's orbital motion is clearly seen, varying between $\pm$15 km/sec over the 3-hr transit, leading for the first time to a determination of the circular velocity of the planet. Combined with the radial velocity variations of the host-star, it results in an assumption-free determination of the mass of the planet and host-star using only Newton's law of gravity (just as for double-line eclipsing binaries).

The detected CO signal seems offset by $\sim$2 km/sec (2$\sigma$) with respect to the systemic velocity of the host star. It suggests the presence of a strong wind flowing from the irradiated dayside to the non-irradiated nightside of the planet within the 0.01-0.1 mbar atmospheric pressure range probed by these observations. Such winds may be driven by the large incident heat flux from the star on the dayside. Indeed three-dimensional circulation models indicate that at low pressure ($<$10 mbar) air should flow from the substellar point towards the antistellar point both along the equator and the poles (Showman et al. 2008). The strength of the carbon monoxide signal suggests a CO mixing ratio of 1-3$\times$10$^{-3}$ in this planet's upper atmosphere. Assuming it is representative for the planet as a whole, and that CO dominates over CH$_4$, it implies that the C/H ratio in HD209458b is a factor 2-6 higher than that of the parent star, and that HD209458b is substantially enriched in heavy elements, to the level of Jupiter and Saturn in our own solar system.

\section{The CRIRES Survey of Hot-Jupiter atmospheres}

\begin{figure}[t]
% \vspace*{-2.0 cm}
\begin{center}
 \includegraphics[width=4.0in]{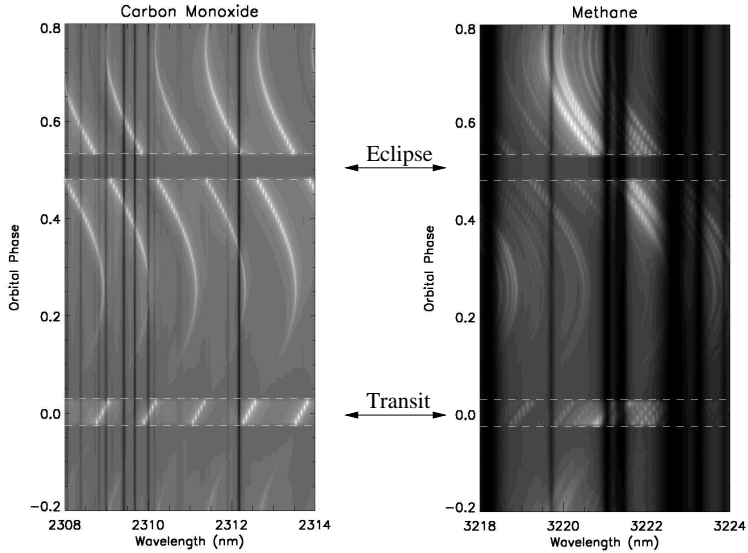} 
% \vspace*{-1.0 cm}
 \caption{Toy model simulations of ground-based spectra of hot Jupiters as function of planet orbital phase, targeting carbon monoxide (left) and methane (right). The vertical bands are telluric absorption, while the planet signals are shown in emission for clarity, assuming uniform emission from the planet dayside, producing a planet signal equally strong as in transmission. Most of the methane transmission features are blocked by the earth atmosphere, but can be observed around phases $\sim$0.4 and $\sim$0.6.}
   \label{fig1}
\end{center}
\end{figure}

The results on carbon monoxide open up a range of very exciting possibilies for new observations with CRIRES, and we have recently been awarded a significant amount of CRIRES observing time to fully explore these. These future observations can be divided into three parts, a) transmission spectroscopy, b) dayside observations of transiting planets, and c) dayside observations of non-transiting planets. 

\noindent {\bf a) Transmission spectroscopy:} A logical next step is to search for a carbon monoxide signal in transmission in the other bright transiting hot Jupiter HD189733b. The CO/CH$_4$ ratio may be less balanced towards CO, because its effective temperature is lower, and its atmospheric scale-height is also smaller. 
But these could well be compensated by the factor 2 higher apparent brightness of the system in K-band, and the deeper transit.

\noindent {\bf b) Dayside spectroscopy of transiting planets:} As is apparent from broad-band spectroscopy (e.g. Swain et al. 2009), dayside absorption features can be just as strong as during transit. Methane and water are expected to produce strong signals in L-band, but are inaccessible with transmission spectroscopy because the strong telluric absorption blocks out most of any possible signal (see Fig.\~1). However, when the dayside emission is probed, the orbital motion of the planet is of great help. When observing at a phase of $\theta \sim$0.4 or $\theta \sim$0.6, we still observe 90\% of the dayside hemisphere, but with the signal blue- or redshifted by 100 km/sec due to the planet's orbital motion - outside the regime where telluric absorption is a problem. This provides some exciting prospects. 

Firstly, such observations will constrain the temperature-pressure profiles of the hot Jupiter atmospheres. Since HD209458b is thought to have a thermal inversion layer, its spectral lines may appear in emission, unambigously proving the inversion to be present. 

Secondly, the uncertainty in radial velocity of our carbon monoxide detection is about 1 km/sec. This means that we can expect to reach a $\sim$1\% uncertainty in the orbital velocity of the planets if we observed them at a red- or blueshift of 100 km/sec, resulting in a mass-determination of the planets and host-stars at a precision of $<$3\%. Not only will these measurements remove any ambiguity in the radii and density of the planets, it will determine these properties of the host stars very accurately, which combined with the effective temperature and metallicity, could result in accurate age determinations for the two systems. 

\noindent {\bf c) Dayside spectroscopy of non-transiting planets:} Since for the dayside spectroscopy it is no longer required that the planets transit their host star, these observations can be extended to non-transiting planets. A great
benefit is that the brightest hot Jupiter systems visible from Paranal, 51 Peg, tau Boo, and HD179949, are 4$-$12 times brighter in K-band than HD209458b. A very exciting prospect for these observations is that a measured planet orbital velocity, in combination with the well-known radial velocity variations of its host-star, will give the planet/star mass ratio, and subsequently the orbital inclination and true mass of the planet. 

\section{Perspectives for the E-ELT}

Our planned observation of methane at 3.2 $\mu$m will fall within the wavelength range of METIS (Brandl et al. 2010), the proposed mid-infrared imager and spectrograph for the European Extremely Large Telescope, and can therefore serve as an interesting test-bed of exoplanet atmosphere observations with the E-ELT. Its baseline design contains a R=100,000 integral field spectrograph ideally suited for this type of work. With its 25x larger collective power, a wealth of new information could be extracted, probing other molecules, structure and variations in absorption features, and/or probing smaller-size planets - possibly down to the super-earth regime.

\acknowledgements

We thank the organisers very much for the extremely stimulating symposium.


\begin{thebibliography}{}

\bibitem[Beaulieu et al.(2009)]{2009arXiv0909.0185B} Beaulieu, J.~P., et 
al.\ 2009, arXiv:0909.0185 
\bibitem[Borucki et al.(2009)]{2009Sci...325..709B} Borucki, W.~J., et al.\ 
2009, Science, 325, 709 
\bibitem[Brandl et al.(2010)]{2010SPIE.7735E..83B} Brandl, B.~R., et al.\ 
2010, SPIE, 7735, pp. 77352G1-77352G-16 
\bibitem[Burrows et al.(2007)]{2007ApJ...668L.171B} Burrows, A., Hubeny, 
I., Budaj, J., Knutson, H.~A., \& Charbonneau, D.\ 2007, ApJL, 668, L171 
\bibitem[Charbonneau et al.(2002)]{2002ApJ...568..377C} Charbonneau, D., 
Brown, T.~M., Noyes, R.~W., \& Gilliland, R.~L.\ 2002, ApJ, 568, 377 
\bibitem[Croll et al.(2010)]{2010ApJ...718..920C} Croll, B., Jayawardhana, 
R., Fortney, J.~J., Lafreni{\`e}re, D., \& Albert, L.\ 2010, ApJ, 718, 920 
\bibitem[de Mooij 
\& Snellen(2009)]{2009A&A...493L..35D} de Mooij, E.~J.~W., \& Snellen, I.~A.~G.\ 2009, A\&A Let, 493, L35 
\bibitem[Grillmair et al.(2008)]{2008Natur.456..767G} Grillmair, C.~J., et 
al.\ 2008, Nature, 456, 767 
\bibitem[Knutson et al.(2007)]{2007Natur.447..183K} Knutson, H.~A., et al.\ 
2007, Nature, 447, 183 
\bibitem[Redfield et al.(2008)]{2008ApJ...673L..87R} Redfield, S., Endl, 
M., Cochran, W.~D., \& Koesterke, L.\ 2008, ApJL, 673, L87 
\bibitem[Sing 
\& L{\'o}pez-Morales(2009)]{2009A&A...493L..31S} Sing, D.~K., \& L{\'o}pez-Morales, M.\ 2009, A\&A, 493, L31 
\bibitem[Showman et al.(2008)]{2008ApJ...682..559S} Showman, A.~P., Cooper, 
C.~S., Fortney, J.~J., \& Marley, M.~S.\ 2008, ApJ, 682, 559 
\bibitem[Sing et al.(2010)]{2010arXiv1008.4795S} Sing, D.~K., et al.\ 2010, 
arXiv:1008.4795 
\bibitem[Snellen et al.(2008)]{2008A&A...487..357S} Snellen, I.~A.~G., 
Albrecht, S., de Mooij, E.~J.~W., \& Le Poole, R.~S.\ 2008, A\&A, 487, 357 
\bibitem[Snellen et al.(2009)]{2009Natur.459..543S} Snellen, I.~A.~G., de 
Mooij, E.~J.~W., \& Albrecht, S.\ 2009, Nature, 459, 543 
\bibitem[Snellen et al.(2010)]{2010Natur.465.1049S} Snellen, I.~A.~G., de 
Kok, R.~J., de Mooij, E.~J.~W., \& Albrecht, S.\ 2010, Nature, 465, 1049 
\bibitem[Swain et al.(2008)]{2008Natur.452..329S} Swain, M.~R., Vasisht, 
G., \& Tinetti, G.\ 2008, Nature, 452, 329 
\bibitem[Swain et al.(2009a)]{2009ApJ...704.1616S} Swain, M.~R., et al.\ 
2009, ApJ, 704, 1616 
\bibitem[Swain et al.(2009b)]{2009ApJ...690L.114S} Swain, M.~R., Vasisht, 
G., Tinetti, G., Bouwman, J., Chen, P., Yung, Y., Deming, D., 
\& Deroo, P.\ 2009, ApJL, 690, L114 
\bibitem[Vidal-Madjar et al.(2003)]{2003Natur.422..143V} Vidal-Madjar, A., 
Lecavelier des Etangs, A., D{\'e}sert, J.-M., Ballester, G.~E., Ferlet, R., 
H{\'e}brard, G., \& Mayor, M.\ 2003, Nature, 422, 143 

\bibitem[Vidal-Madjar et al.(2004)]{2004ApJ...604L..69V} Vidal-Madjar, A., 
et al.\ 2004, ApJL, 604, L69 
\end{thebibliography}
\end{document}